\documentclass[a4paper]{article}

\pdfoutput=1 

\usepackage[T1]{fontenc}
\usepackage[utf8]{inputenc}
\usepackage[english]{babel}
\usepackage{graphicx}
\usepackage{epsfig}
\usepackage{amsmath}
\usepackage{amssymb}
\usepackage{amsthm}
\usepackage{mathtools}
\usepackage{url}
\usepackage[breaklinks]{hyperref}
\usepackage{float}
\usepackage{pifont}
\usepackage[table]{xcolor}
\usepackage{lmodern}
\usepackage{xspace}
\usepackage{bbold}
\usepackage[capitalize,nameinlink]{cleveref}
\usepackage{breakcites}
\usepackage{framed}
\usepackage{setspace}

\newcommand{\T}{\intercal}
\DeclareMathOperator{\sinc}{sinc}
\DeclareMathOperator{\R}{\mathbb{R}}

\DeclareMathOperator{\e}{e}
\DeclareMathOperator{\D}{\mathrm{d}\!}

\author{Christian Himpe%
\thanks{ORCiD: \href{http://orcid.org/0000-0003-2194-6754}{0000-0003-2194-6754}, %
Contact: \href{mailto:christian.himpe@wwu.de}{\nolinkurl{christian.himpe@wwu.de}}, %
Institute for Numerical Analysis, University of M\"unster, Orleans-Ring~10, 48149 M\"unster, Germany}%
}

\title{\texttt{emgr} -- EMpirical GRamian Framework \\ Version~5.99 \\[1ex] \large Software Release Paper}

\newtheoremstyle{thm}{\topsep}{\topsep}{\normalfont \itshape}{}{\normalfont \bfseries}{}{\newline}{}
\theoremstyle{thm}

\newcounter{dummy}
\newtheorem{mydefine}[dummy]{Definition}

\date{}

\begin{document}

\maketitle

\begin{abstract}
Version~5.99 of the empirical Gramian framework -- \texttt{emgr} -- completes a
development cycle which focused on parametric model order reduction of gas
network models while preserving compatibility to the previous development for
the application of combined state and parameter reduction for neuroscience
network models. Secondarily, new features concerning empirical Gramian types,
perturbation design, and trajectory post-processing, as well as a Python version
in addition to the default MATLAB / Octave implementation, have been added. This 
work summarizes these changes, particularly since \texttt{emgr} version~5.4,
see \textsc{Himpe}, 2018 [Algorithms~11(7):~91],
and gives recent as well as future applications, such as parameter
identification in systems biology, based on the current feature set.
\end{abstract}




\section{Project History and Overview}
The empirical Gramian framework (\texttt{emgr}) is an open-source MATLAB (and Octave-compatible) software package for the computation of
empirical system Gramians and empirical covariance matrices,
which are (approximations to the) essential operators in (nonlinear) system theory.

Originally, \texttt{emgr} was started to provide reusable computational kernels with a unified interface,
while the interface is inspired by the \texttt{gram} function of the Control System Toolbox~\cite{consystbx}
and the ``Nonlinear Model Reduction Routines''~\cite{modredrou}.

The first release (version~0.9) coincided with the ``MoRePaS 2'' workshop\footnote{see: \url{https://web.archive.org/web/20121219154629/http://www.morepas.org:80/workshop2012/index.html}} in 2012,
while a first summary of the capabilities is published in~\cite{morHimO13} (based on version~1.3),
and detailed descriptions of \texttt{emgr} are given in~\cite{morHim17} (based on version~3.9) and~\cite{morHim18b} (based on version~5.4).
Marking the ten-year anniversary of \texttt{emgr}'s development, version~5.99~\cite{morHim22} was released.
And relative to version~5.4, various major features have been added, which this work summarizes.

The (empirical) system Gramian matrices have a multitude of system theoretic applications,
which include model reduction, parameter identification, control configuration selection,
sensitivity analysis, optimal placement, nonlinearity quantification, or system characterization via indices and invariants.
Beyond system and control theory, areas such as uncertainty quantification use numerically approximations of operators
which are computable as empirical Gramians, for example Hessians~\cite{morLieFWetal13}.
Recent interesting uses for system Gramians in particular applications are:
pose detection~\cite{AvaM19}, traffic networks~\cite{BiaP20}, tau functions~\cite{BloN21},
and vulnerability analysis~\cite{Bab22}.

\section{New Features}
A detailed description of the features up to and including \texttt{emgr}~5.4 is given in~\cite{morHim18b}.
This section summarizes the major new features implemented since version~5.4 onwards until the latest version~5.99.
These features are grouped in to five categories: Gramian variants, input functions, trajectory weighting, parameter identifiability, and Python version.

\subsection{Gramian Variants}
\texttt{emgr} provides seven empirical Gramians: controllability, observability, cross, linear cross, sensitivity, identifiability and joint Gramian.
Of those, only the cross-Gramian derived empirical Gramians (cross, linear cross and joint Gramian) used to provide a variant,
specifically for non-square or non-symmetric systems~\cite[Sec.~3.1.5]{morHim18b}.
Over the recent releases, variants also for the controllability- and observability-based Gramians were added.

It is noted here, that the loadability Gramian from \cite{morTolA17} is not a system Gramian but a standard Gram matrix \cite{wikigram},
and hence does not need to be computed via \texttt{emgr}.

\subsubsection{Output Controllability Gramian}
The output reachability Gramian, or more generally the output controllability Gramian \cite{KreS64},
encodes the controllability of the output $y : \R \to \R^Q$ instead of the controllability of the state $x : \R \to \R^N$.
It has various applications in system theory, for example in control configuration selection~\cite{Hal08},
while more recently the empirical output-controllability covariance matrix (EOCCM)~\cite{MenO21} is employed for parameter identifiability.
An empirical output controllability Gramian can always be computed via the empirical controllability Gramian,
given a linear output operator~$C$ of the underlying system,
\begin{align*}
 \widehat{W}_{OC} = C \widehat{W}_C C^\T.
\end{align*}
However in version~5.8, direct computation of the output controllability Gramian was included;
not only to provide a more memory efficient computation for large-scale systems
by computing an empirical output controllability Gramian from output trajectory data directly, instead of state trajectories,
but also to approximate the output controllability for systems with nonlinear output operators.
Following the format of~\cite{morHim18b}, it is defined as:

\pagebreak

\begin{mydefine}[Empirical Output Controllability Gramian]
Given non-empty sets $E_u$ and $S_u$, the \textbf{empirical output controllability Gramian} $\widehat{W}_{OC} \in \R^{Q \times Q}$ is defined as:
\begin{align*}
 \widehat{W}_{OC} &:= \frac{1}{|S_u|} \sum_{k=1}^{|S_u|} \sum_{m=1}^M \frac{1}{c_k^2} \int_0^T \Psi^{km}(t) \D t \\
 \Psi^{km}(t) &= (y^{km}(t) - \bar{y}^{km}) (y^{km}(t) - \bar{y}^{km})^\T \in \R^{Q \times Q},
\end{align*}
with the output trajectories $y^{km}(t) \in \R^Q$ for the input configurations \linebreak \mbox{$\hat{u}^{km}(t) = c_k e^m \circ u(t) + \bar{u}$}, $c_k \in S_u$, $e_m \in E_u$,
and offsets $\bar{u} \in \R^M$, $\bar{x}^{km} \in \R^N$.
\end{mydefine}

\subsubsection{Average Observability Gramian}
An idea similar to the non-symmetric (empirical) cross Gramian \cite[Sec.~3.1.5]{morHim18b},
is an average observability Gramian, which is hinted at in \cite{RonM16}.
Practically, this means for multiple output systems, that all outputs are summed up yielding a single (average) output.
This variant, added in version~5.7, also extends to the augmented empirical observability Gramian,
and thus the empirical identifiability Gramian.

The empirical local observability Gramian \cite{KreI09}, could have been a
potential variant, yet, \cite[Sec.~II.A]{RonM16} illustrates why the standard empirical observability Gramian suffices.

\subsection{Input Functions}
Already up to version~5.4, \texttt{emgr} provided means to pass a custom function of a single (time) argument as input function
for the empirical Gramian computation or select from the included default input functions:
impulse, decaying chirp, or pseudo-random binary sequence. 
Subsequently, two more default input functions were implemented:

\subsubsection{Step Function Input}
Since for (semi-discrete) hyperbolic partial differential equation models, with inputs and outputs at the boundaries, step functions
are a relevant training input, which was demonstrated heuristically in~\cite{morGruHS19},
in version~5.7 a constant ``step'' function was added as default input function:
\begin{align*}
 u_{\text{step}}(t) := 1.
\end{align*}
This training input became the default for data-driven reduced order gas network models in the \texttt{morgen} platform~\cite{morHimGB21a},
which utilizes \texttt{emgr} as model reduction back-end.

\subsubsection{Sine Cardinale Input}
A smooth alternative to impulse input is a \emph{sine cardinale} (sinc) input, as it was employed in \cite{ArjCH11}. \pagebreak
In version~5.8, a scaled $\sinc$ input function has been included into the set of default inputs:
\begin{align*}
 u_{\text{sinc}}(t) := \begin{cases} \frac{\sin(t h^{-1})}{t h^{-1}} & t \neq 0, \\ ~~~~~1 & t = 0, \end{cases}
\end{align*}
for time-step width $h>0$.

\subsection{Trajectory Weighting}
As mentioned in \cite[Sec.~5.3]{morHim18b} trajectory weighting, could be implemented by using the custom inner product interface.
However, a set of weighting functions has been included, originally motivated by time domain weighting.
Nonetheless, trajectory independent weightings like in \cite{Mit69} are still achieved via the inner product interface.

\subsubsection{Time-Weighting}
Time-domain weighting of Gramians was initially proposed in~\cite{morSchD95}, particularly, using monomials of the time variable,
and also provides an error bound~\cite{morSre02} if used in conjunction with balanced truncation \cite{morBreS21}.  
In version~5.8, linear and a quadratic time-domain weighting was included, based on a time-weighted linear system Gramian and \cite[Sec.~3.1]{morHim18b}: 
\begin{align*}
 W_* &= \frac{1}{r!} \int_0^\infty t^r \e^{A_1 x(t)} D \e^{A_2 x(t)} \D t \\
 \to \widehat{W}_* &= \frac{1}{\dots} \sum \dots \sum \frac{1}{\dots} \; \frac{1}{r!} \int t^r \; \Psi(t) \D t
\end{align*}
for $r \in \{1,2\}$ and all computable empirical Gramians.
Practically, such time-weighting emphasizes ``later'' parts of a simulated trajectory in the empirical Gramian,
over the ``earlier'' parts, like the initial state / output.
Note, that the scaling factor $\frac{1}{r!}$ from~\cite{morSre02} is included for convenience,
in case for a typical use in conjunction with balanced truncation.

\subsubsection{Reciprocal Time-Weighting}
Furthermore, in the latest version~5.99 a time-reciprocal weighting, also based on a weighted linear system Gramian and~\cite[Sec.~3.1]{morHim18b}:
\begin{align*}
 W_* &=  \int_0^\infty \frac{1}{\sqrt{\pi t}} \e^{A_1 x(t)} D \e^{A_2 x(t)} \D t \\
 \to \widehat{W}_* &= \frac{1}{\dots} \sum \dots \sum \frac{1}{\dots} \int \; \frac{1}{\sqrt{\pi t}} \Psi(t) \D t
\end{align*}
following~\cite[Sec.~3.2]{morGlo87} was included, i.e. to allow numerical verification of the lower error bound presented in there,
and to provide a time-weighting emphasizing ``earlier'' parts of a simulated trajectory, in contrast to the typical time-weighting.
Note, that practically at time $t=0$ the scaling factor is set to~$\sqrt{\frac{2}{\pi \Delta t}}$,
which was determined heuristically to be suitable.

\newpage

\subsubsection{Column-Based Weighting}
The column-based weighting originates in an approach to suboptimal control from \cite[Def.~2]{HyuMV17},
and defines a weighted Gramian which normalizes the state $x$ (or output $y$) at each time instance by its length, i.e.:
\begin{align*}
 \hat{x}(t) &:= x(t) \cdot \|x(t)\|_2^{-1}, \\
 \hat{y}(t) &:= y(t) \cdot \|y(t)\|_2^{-1}.
\end{align*}

\subsubsection{Row-Based Weighting}
The row-based weighting is based on component-wise scale normalization.
This means each component $i$ of the utilized state $x$ or output $y$ trajectories for the Gramians 
is normalized by its (absolute) maximum value, i.e.:
\begin{align*}
 \hat{y}_i(t) &:= x_i(t) \cdot \|x_i\|_\infty^{-1}, \\
 \hat{y}_i(t) &:= y_i(t) \cdot \|y_i\|_\infty^{-1},
\end{align*}
which means all component (output) trajectories evolve in the interval $[-1,1]$.

\subsection{Parameter Identifiability}
More recently parameter identification of nonlinear systems with low-dimensional state-space,
but high-dimensional parameter-space became a use-case for \linebreak \texttt{emgr},~\cite{FalHKetal22}.
This motivated the following enhancements:

\subsubsection{Schur Complement}
Initially, parameter identification was the basis for the combined state and parameter reduction~\cite{morHim17}
of nonlinear systems with high-dimensional state and parameter spaces,
but homogeneous parameters,
hence the matrix-inverse inside the Schur complement is only roughly approximated by a truncated Neumann series.

To improve accuracy of the empirical (cross-)identifiability Gramian,
a more accurate Schur complement option was added in version~5.99, which computes the
inner inverse as Moore-Penrose pseudo-inverse:
\begin{align*}
 W_{I,\text{exact}} &= W_P - W_M^\T W_O^+ W_M, \\
 W_{\ddot{I},\text{exact}} &= -\frac{1}{2} W_m^\T (W_X + W_X^\T)^+ W_m.
\end{align*}

\subsubsection{Parameter Centering}
Originally parameter-related Gramians (empirical sensitivity, identifiability, joint Gramian)
required a minimum and maximum parameter to define the range of perturbation.
Recently in version~5.99, a mode was added, which requires minimum, maximum and nominal parameter,
which then sets up a range of perturbation with respect to the nominal value instead of the minimum or (logarithmic) mean.

\subsection{Python Version}
Since version~5.6, a Python~(version~3) variant of \texttt{emgr} is also maintained\footnote{see:~\texttt{py/emgr.py}},
that provides the same features, and closely resembles the MATLAB interface and function signature.
The elaborate testing, prototype and wrapper code, as for the MATLAB variant, is not available yet.
However, a combinatorial testing of configurations is supported\footnote{see:~\texttt{py/emgrtest.py}}.

\section{Application Demonstration}
A current application for \texttt{emgr}'s empirical Gramians is,
after combined state and parameter reduction for brain connectivity inference~\cite{morHim17},
and parametric model order reduction for gas networks~\cite{morHimGB21a},
parameter identification for systems biology models.

Such application is exemplarily demonstrated on a benchmark model --
the IL13-Induced JAK/STAT signaling model from \cite{RauKSetal14},
which is also tested in \cite{VilBP16} and \cite{StiJ21}.
This model has $\dim(u(t)) = 1$ input, $\dim(x(t)) = 10$ states, $\dim(y(t)) = 8$ outputs and $\dim(\theta) = 23$ parameters,
and the following nonlinear vector field as well as linear output function (in simplified form):
{\setstretch{1.3}
\begin{align*}
 \dot{x}(t) &= \begin{bmatrix}
                \theta_6 x_2(t) - \theta_5 x_1 - c_1 \theta_1 x_1(t) u_1(t) \\
                \theta_5 x_1(t) - \theta_6 x_2(t) \\
                \theta_2 x_3(t) (x_6(t) - c_3) + c_1 \theta_1 x_1(t) u_1(t) \\
                -\theta_3 x_4(t) - \theta_2 x_3(t) (x_6(t) - c_3) \\
                \theta_3 x_4(t) - \theta_4 x_5(t) \\
                -c_2 \, \theta_8 (x_6(t)- c_3) - \frac{\theta_7 x_3(t) x_6(t)}{\theta_{13} x_1(t) + 1} - \frac{\theta_7 x_4(t) x_6(t)}{\theta_{13} x_1(t) + 1} \\
                \theta_9 x_7(t) (x_6(t) - c_3) - c_2 x_{10}(t) (x_7(t) - c_4) \\
                -\theta_{11} (x_7(t) - c_4) \\
                -c_1 \theta_{12} x_9(t) u_1(t) \\
                \frac{\theta_{14} x_8(t)}{\theta_{15} + x_8(t)} - \theta_{16} x_{10}(t)
               \end{bmatrix}, \\
       y(t) &= \begin{bmatrix}
                x_1(t) + x_3(t) + x_4(t) \\
                \theta_{18} (x_3(t) + x_4(t) + x_5(t) + c_5 - x_9) \\
                \theta_{19} (x_4(t) + x_5(t)) \\
                \theta_{20} (c_3 - x_6(t)) \\
                \theta_{21} x_8(t) \\
                \theta_{17} \theta_{22} \theta_{11}^{-1} x_8(t) \\
                x_{10}(t) \\
                c_4 - x_7(t)
               \end{bmatrix},
\end{align*}}
with constants $c_1 = 2.265$, $c_2 = 91$, $c_3 = 2.8$, $c_4 = 165$, and $c_5 = 0.34$.
Additionally, the initial state has a parameter dependency:
\begin{align*}
 x_0 = \begin{bmatrix} 1.3 & \theta_{23} & 0 & 0 & 0 & c_3 & c_4 & 0 & c_5 & 0 \end{bmatrix}^\T.
\end{align*}
As initial-state-parameters are not supported by default in \texttt{emgr}, but can be
emulated by providing a solver wrapper\cite[Sec.~5.5]{morHim18b} which sets the parameter in the initial states
and passes the updated initial state to the actual solver.

To assess the parameter identifiability, the empirical identifiability Gramian
is computed via the augmented empirical observability Gramian,
of which its singular value decomposition $W_I = U \Sigma U^\T$ is analyzed.
This numerical experiment is conducted in MATLAB~2022a on a AMD~Ryzen~5~4500U with 16GiB~RAM.

\begin{figure}\centering
 \includegraphics[width=.95\textwidth]{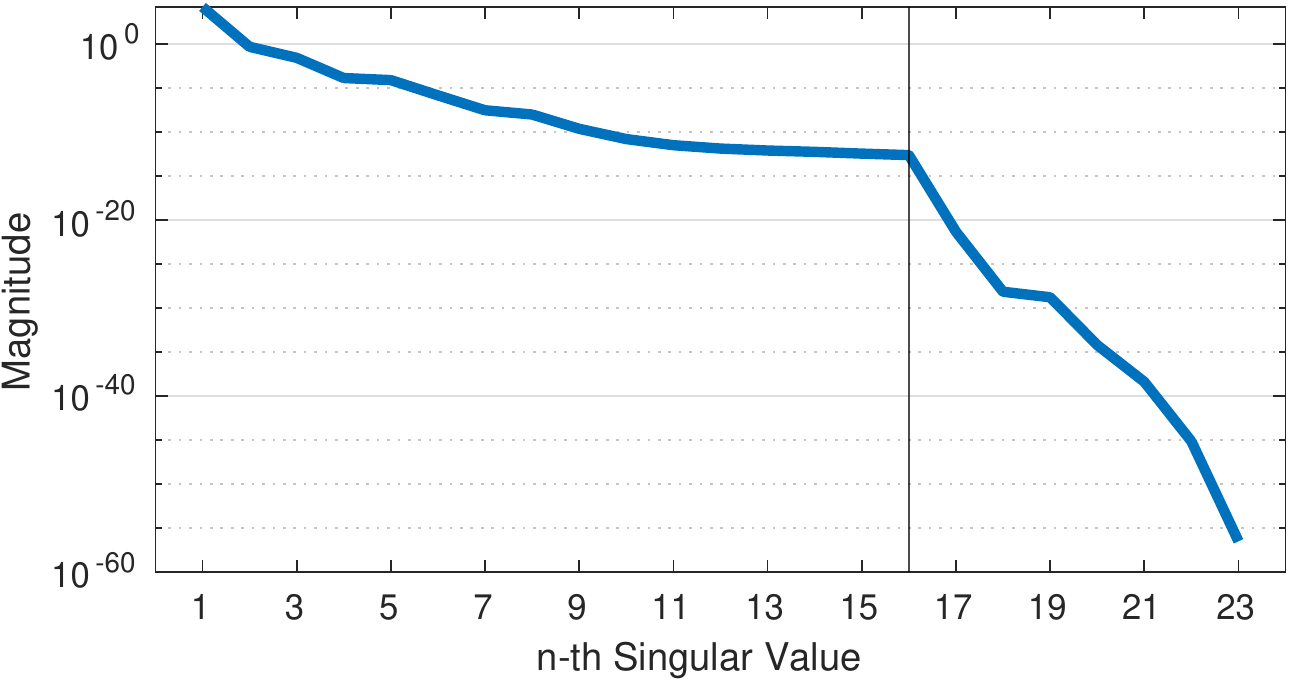}
 \caption{Singular values of the empirical identifiability Gramian for the JAK/STAT benchmark system's parameters.}
 \label{fig:singval}
\end{figure}

The singular values $\Sigma_{ii}$ are plotted in \cref{fig:singval} and indicate that the singular vectors,
associated to the seven smallest singular values, $\{ u_{17}, \dots, u_{23} \}$ contain linear combinations
of the original parameters that are least identifiable. To reconstruct the contributions of those,
these singular vectors are summed after taking their element-wise absolute value, $\bar{u}_2 := \sum_{k=17}^{23} |u_k|$,
which yields the overall contribution of original parameter fractions to the ``unidentifiable'' singular vectors.
The dominant contribution is given by the five structurally unidentifiable parameters
$\{ \theta_{11}, \theta_{15}, \theta_{17}, \theta_{21}, \theta_{22} \}$, as well as by the
practically unidentifiable parameters $\{\theta_4, \theta_{14}, \theta_{16}, \theta_{19} \}$.
The remaining practically unidentifiable parameters $\{\theta_3 , \theta_{12} \}$ are not contributing.
However, all structurally unidentifiable parameters are located, and particularly, all identifiable parameters
do not contribute (the respective elements of $\bar{u}_2$ have relative magnitudes below $10^{-7}$)
to the singular vectors of small singular values.
Thus, the results of the empirical-Gramian-based parameter identification agrees
with other studies on this system in terms structural identifiability.
Furthermore, the matrix of singular vectors associated to the dominant singular values,
represents a lower-dimensional reparametrization of the system.

\section{Summary}

In a decade of research in empirical system Gramians and working on \texttt{emgr},
I conclude that an empirical-Gramian-based approach typically gives an acceptable approximate answer,
no matter the system's complexities, which makes this data-driven mathematical technology
a somewhat universal tool for linear and nonlinear control and system-theory and engineering.
Lastly, I note that more information and documentation on \texttt{emgr} can be found on:

\begin{center}
 \url{https://gramian.de}
\end{center}

\subsection*{Code Availability}
\begin{minipage}{\linewidth} 
  \begin{framed}
    The source code of the numerical experiments is licensed under \textsc{BSD-2-Clause License},
    can be obtained from:
    \begin{center}
      \href{http://doi.org/10.5281/zenodo.7048585}{\texttt{doi:10.5281/zenodo.7048585}}
    \end{center}
    and is authored by: \textsc{C.~Himpe}.
  \end{framed}
\end{minipage}

\bibliographystyle{apalike}
\bibliography{mor,csc,software,extra}

\begin{thebibliography}{}

\bibitem[Arjona et~al., 2011]{ArjCH11}
Arjona, M.~A., Cisneros-Gonaz{\'a}lez, M., and Hern{\'a}ndez, C. (2011).
\newblock Parameter estimation of a synchronous generator using a sine cardinal
  perturbation and mixed stochastic-deterministic algorithms.
\newblock {\em IEEE Transactions on Industrial Electronics}, 58(2):486--493.

\bibitem[Avant and Morganson, 2019]{AvaM19}
Avant, T. and Morganson, K.~A. (2019).
\newblock Observability properties of object pose estimation.
\newblock In {\em Proceedings of the American Control Conference}, pages
  5134--5140.

\bibitem[Babazadeh, 2022]{Bab22}
Babazadeh, M. (2022).
\newblock Gramian-based vulnerability analysis of dynamic networks.
\newblock {\em IET Control Theory \& Application}, 16(6):625--637.

\bibitem[Bianchin and Pasqualetti, 2020]{BiaP20}
Bianchin, G. and Pasqualetti, F. (2020).
\newblock {G}ramian-based optimization for the analysis and control of traffic
  networks.
\newblock {\em IEEE Transactions on Intelligent Transportation Systems},
  21(7):3013--3024.

\bibitem[Blower and Newsham, 2021]{BloN21}
Blower, G. and Newsham, S.~L. (2021).
\newblock Tau functions associated with linear systems.
\newblock In {\em Operator Theory, Functional Analysis and Applications}, pages
  63--94.

\bibitem[Breiten and Stykel, 2021]{morBreS21}
Breiten, T. and Stykel, T. (2021).
\newblock Balancing-related model reduction methods.
\newblock In Benner, P., Grivet-Talocia, S., Quarteroni, A., Rozza, G.,
  Schilders, W., and Silveira, L.~M., editors, {\em System- and Data-Driven
  Methods and Algorithms}, volume~1 of {\em Model Order Reduction}, chapter~2,
  pages 15--56. De Gruyter.

\bibitem[Falkenhagen et~al., 2022]{FalHKetal22}
Falkenhagen, U., Himpe, C., Kloft, C., Knoechel, J., and Huisinga, W. (2022).
\newblock Sample-based robust model order reduction for nonlinear systems
  biology models.
\newblock In {\em In Preparation}.

\bibitem[Glover, 1987]{morGlo87}
Glover, K. (1987).
\newblock Model reduction: A tutorial on {H}ankel-norm methods and lower bounds
  on {$L^2$}~errors.
\newblock {\em IFAC Proceedings Volume (10th Triennial IFAC Congress on
  Automatic Control)}, 20(5):293--298.

\bibitem[Grundel et~al., 2019]{morGruHS19}
Grundel, S., Himpe, C., and Saak, J. (2019).
\newblock On empirical system {G}ramians.
\newblock {\em Proc. Appl. Math. Mech.}, 19(1):e201900006.

\bibitem[Halvarsson, 2008]{Hal08}
Halvarsson, B. (2008).
\newblock Comparison of some {G}ramian based interaction measures.
\newblock In {\em 2008 IEEE Int Symposium on Computer-Aided Control System
  Design}, pages 128--143.

\bibitem[Himpe, 2017]{morHim17}
Himpe, C. (2017).
\newblock {\em Combined State and Parameter Reduction for Nonlinear Systems
  with an Application in Neuroscience}.
\newblock PhD thesis, Westf{\"a}lische Wilhelms-Universit{\"a}t M{\"u}nster.
\newblock Sierke Verlag G{\"o}ttingen, ISBN 9783868448818.

\bibitem[Himpe, 2018]{morHim18b}
Himpe, C. (2018).
\newblock emgr -- the {E}mpirical {G}ramian {F}ramework.
\newblock {\em Algorithms}, 11(7):91.

\bibitem[Himpe, 2022]{morHim22}
Himpe, C. (2022).
\newblock {emgr -- EMpirical GRamian} framework (version~5.99).
\newblock \url{https://gramian.de}.

\bibitem[Himpe et~al., 2021]{morHimGB21a}
Himpe, C., Grundel, S., and Benner, P. (2021).
\newblock Model order reduction for gas and energy networks.
\newblock {\em Journal of Mathematics in Industry}, 11:13.

\bibitem[Himpe and Ohlberger, 2013]{morHimO13}
Himpe, C. and Ohlberger, M. (2013).
\newblock A unified software framework for empirical {G}ramians.
\newblock {\em J. Math.}, 2013:1--6.

\bibitem[Hyun et~al., 2017]{HyuMV17}
Hyun, N.-s.~P., Murali, V., and Verriest, E.~I. (2017).
\newblock Minimum sensitivity analysis for accurate open-loop controllers in
  linear systems using weighted {G}ramians.
\newblock In {\em Proceedings of the IEEE 56th Annual Conference on Decision
  and Control}, pages 114--119.

\bibitem[Kreindler and Sarachik, 1964]{KreS64}
Kreindler, E. and Sarachik, P.~E. (1964).
\newblock On the concepts of controllability and observability of linear
  systems.
\newblock {\em {IEEE} Trans. Autom. Control}, 9(2):129--136.

\bibitem[Krener and Ide, 2009]{KreI09}
Krener, A. and Ide, K. (2009).
\newblock Measures of unobservability.
\newblock In {\em Proceedings of the 48th IEEE Conference on Decision and
  Control, 2009 held jointly with the 2009 28th Chinese Control Conference},
  pages 6401--6406.

\bibitem[Lieberman et~al., 2013]{morLieFWetal13}
Lieberman, C.~E., Fidkowski, K., Willcox, K., and Van Bloemen~Waanders, B.
  (2013).
\newblock Hessian-based model reduction: large-scale inversion and prediction.
\newblock {\em Int. J. Numer. Methods Fluids}, 71(2):135--150.

\bibitem[MathWorks, nd]{consystbx}
MathWorks ({n.d.}).
\newblock {Control System Toolbox}.

\bibitem[M{\'e}ndez-Blanco and {\"O}zkan, 2021]{MenO21}
M{\'e}ndez-Blanco, C.~S. and {\"O}zkan, L. (2021).
\newblock Local parameter identifiability of large-scale nonlinear models based
  on the output sensitivity covariance matrix.
\newblock {\em IFAC-PapersOnLine (16th IFAC Symposium on Advanced Control of
  Chemical Processes ADCHEM 2021)}, 54(3):415--420.

\bibitem[Mitra, 1969]{Mit69}
Mitra, D. (1969).
\newblock {$W$} matrix and the geometry of model equivalence and reduction.
\newblock {\em Proceedings of the Institution of Electrical Engineers},
  116(6):1101--1106.

\bibitem[Raue et~al., 2014]{RauKSetal14}
Raue, A., Karlsson, J., Saccomani, M.~P., and Jirstrand, M. (2014).
\newblock Comparison of approaches for parameter identifiability analysis of
  biological systems.
\newblock {\em Bioinformatics}, 30(10):1440--1448.

\bibitem[Rong and Michael, 2016]{RonM16}
Rong, Z. and Michael, N. (2016).
\newblock Detection and prediction of near-term state estimation degradation
  via online nonlinear observability analysis.
\newblock In {\em IEEE International Symposium on Safety, Security, and Rescue
  Robotics (SSRR)}, pages 28--33.

\bibitem[Schelfhout and De~Moor, 1995]{morSchD95}
Schelfhout, G. and De~Moor, B. (1995).
\newblock Time-domain weighted balanced truncation.
\newblock In {\em 3rd European Control Conference}, pages 1--4.

\bibitem[Sreeram, 2002]{morSre02}
Sreeram, V. (2002).
\newblock Frequency response error bounds for time-weighted balanced
  truncation.
\newblock In {\em Proceedings of the 41st IEEE Conference on Decision and
  Control}, pages 3330--3331.

\bibitem[Stigter and Joubert, 2021]{StiJ21}
Stigter, J.~D. and Joubert, D. (2021).
\newblock Computing measures of identifiability, observability and
  controllability for a dynamic system model with the {StrucID} app.
\newblock {\em IFAC PapersOnLine (19th IFAC Symposium on System
  Identification)}, 54(7):138--143.

\bibitem[Sun and Hahn, nd]{modredrou}
Sun, C. and Hahn, J. ({n.d.}).
\newblock {Nonlinear Model Reduction Routines for MATLAB}.

\bibitem[Tolks and Ament, 2017]{morTolA17}
Tolks, C. and Ament, C. (2017).
\newblock Model order reduction of glucose-insulin homeostasis using empirical
  {G}ramians and balanced truncation.
\newblock {\em IFAC-PapersOnline (Proceedings of the 20th IFAC World
  Congress)}, 50(1):14735--14740.

\bibitem[Villaverde et~al., 2016]{VilBP16}
Villaverde, A.~F., Barreiro, A., and Papachristodoulou, A. (2016).
\newblock Structural identifiability of dynamic systems biology models.
\newblock {\em PLOS Computational Biology}, 12(10):e1005153.

\bibitem[{Wikipedia contributors}, 2022]{wikigram}
{Wikipedia contributors} (2022).
\newblock Gram matrix --- {Wikipedia}{,} the free encyclopedia.
\newblock
  \url{https://en.wikipedia.org/w/index.php?title=Gram_matrix&oldid=1093332712}.
\newblock [Online; accessed 23-August-2022].

\end{thebibliography}

\end{document}